\documentclass{vgtc}                          % final (conference style)
\graphicspath{{figures/}{pictures/}{images/}{./}} % where to search for the images

\usepackage{times}                     % we use Times as the main font
\usepackage{mathptmx}                  % use matching math font

\usepackage{cite}

\onlineid{0}

\vgtccategory{Research}

\vgtcinsertpkg

\title{ReVoicer: Conversational Voice Annotation for \\Human-Centered, LLM-Assisted Peer Review}

\author{Matt Gottsacker\thanks{e-mail: mattg@ucf.edu}\\
\scriptsize SREAL, University of Central Florida
\and Ahinya Alwin\thanks{e-mail: ahinya@ucf.edu}\\
\scriptsize SREAL, University of Central Florida
\and Hiroshi Furuya\thanks{e-mail: furuya@ucf.edu}\\
\scriptsize SREAL, University of Central Florida
\and Robert W. Lindeman\thanks{e-mail: rob.lindeman@canterbury.ac.nz}\\
\scriptsize HIT Lab NZ, University of Canterbury
\and Gerd Bruder\thanks{e-mail: bruder@ucf.edu}\\
\scriptsize SREAL, University of Central Florida
\and Gregory F. Welch\thanks{e-mail: welch@ucf.edu}\\
\scriptsize SREAL, University of Central Florida
}
\teaser{
  \centering
  \vspace{-1ex}
  \includegraphics[width=\linewidth]{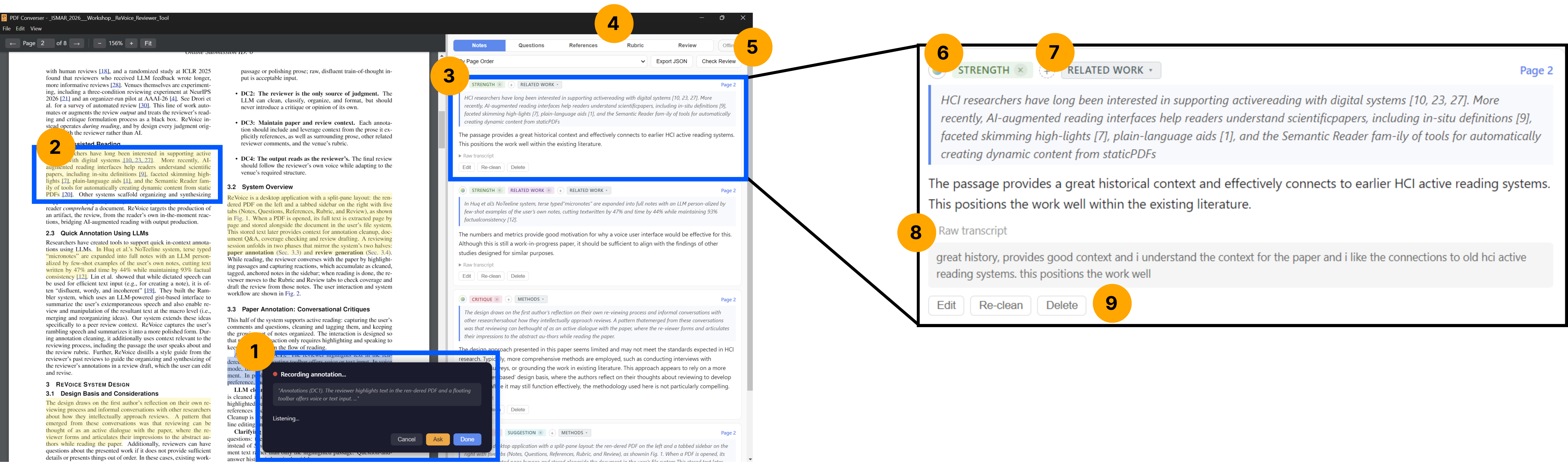}
  \vspace{-3ex}
  \caption{Annotated screenshot of ReVoicer. Left: the PDF viewer with color-coded highlights and the voice recording overlay (1), from which the reviewer submits the spoken comment as an annotation (Done) or as a question to the paper (Ask), and previously annotated passages (2). Center: the sidebar's Notes tab with the cleaned, tagged note cards (3), tabs for the other parts of the review workflow: Q\&A history, references, rubric, and review generation (4), and actions for exporting annotations and running the rubric coverage check (5). Right: an enlarged note card showing the comment type tag (6), the paper section tag (7), the expandable raw transcript of the original speech (8), and per-note controls to edit, re-clean, or delete (9).}
  \label{fig:teaser}
}

\abstract{
We present \textit{ReVoicer}, a prototype system that supports peer reviewers by letting them converse with a paper as they read it.
The reviewer highlights a passage and speaks (or types) a train-of-thought comment. A large language model then cleans the comment using the surrounding prose as context, tags it by comment type, and anchors it to the passage.
After the reviewer finishes reading, \textit{ReVoicer} checks the accumulated notes against a venue-specific rubric and reports coverage gaps to assist with further reflection.
Then \textit{ReVoicer} drafts a review composed from the reviewer's comments, written to a style guide distilled from the reviewer's past reviews.
We describe the system's design rationale and implementation, and we outline plans for future evaluations.
} % end of abstract


\usepackage{tikz}
\usetikzlibrary{positioning, arrows.meta, backgrounds, fit, calc}
\definecolor{rvHuman}{HTML}{4477AA}
\definecolor{rvLLM}{HTML}{EE7733}

\usepackage{listings}
\lstdefinestyle{prompt}{
  breaklines=true,
  basicstyle=\scriptsize\rmfamily,
  columns=fullflexible,
  keepspaces=true,
  frame=single,
  framesep=4pt,
  xleftmargin=2pt,
  xrightmargin=2pt,
  aboveskip=4pt,
  belowskip=8pt,
  literate={–}{{--}}1 {—}{{---}}1 {’}{{'}}1
}

\begin{document}


\firstsection{Introduction}

\maketitle

%% Introduction body. The section heading is issued by \firstsection{Introduction} in 00_manuscript.tex (vgtc convention).

Peer review is essential labor for the scientific research community, and the need for high-quality reviews is growing faster than the pool of qualified reviewers across fields.
The scientific literature has grown at an accelerating rate for decades, with global publication output doubling roughly every 17 years~\cite{bornmann2021growth}, straining reviewers across disciplines~\cite{shah2022challenges}.
Recruiting reviewers has also become harder over time. One study found that at four out of six ecology and evolution journals, the percentage of review invitations that led to a review fell from roughly 56\% to 37\% between 2003 and 2015~\cite{fox2017recruitment}.
The strain is especially acute at large computing venues. The flagship human-computer interaction conference, ACM CHI, received 5{,}014 complete submissions in 2025, up 58\% from 2023, prompting a new assisted desk-rejection process to manage reviewer load~\cite{chi2025report,chi2026adr}.
One of the largest artificial intelligence (AI) conferences, AAAI, received almost 23{,}000 full paper submissions in 2026~\cite{biswas2026aaai}.
The reviews produced under this load are also noisy, even at well-resourced venues. In a 2014 review consistency experiment at the machine learning conference NeurIPS, two independent committees disagreed on about 25\% of decisions~\cite{cortes2021inconsistency}, a result replicated at NeurIPS 2021 with 23\% disagreement~\cite{beygelzimer2023arbitrary}. In both examples, roughly half of the accepted-paper list would change if the process were rerun~\cite{cortes2021inconsistency,beygelzimer2023arbitrary}.
Recent position papers deem this situation a ``peer review crisis'' that threatens the reputation of the entire established peer review process~\cite{kim2025position}.
In short, researchers across fields face more submissions, too few qualified reviewers, and high variance in review quality and outcomes.

One response to this crisis has been to apply large language models (LLMs) to generate reviews.
Researchers have found that LLM-generated feedback overlaps substantially with human reviews~\cite{liang2024can}.
Additionally, a large randomized study at the machine learning conference ICLR in 2025 found that LLM feedback measurably improved human reviews~\cite{thakkar_largescale_2026}, and some of the largest AI conferences, including NeurIPS and AAAI, are piloting AI assistance in the review process at scale~\cite{neurips2026experiment,biswas2026aaai}.
% Yet full automation raises persistent concerns: hallucinated or unverifiable critiques, loss of human oversight and accountability, homogenized feedback, unclear authorship of LLM-produced text (flagged as an open concern even in LLM writing tools~\cite{lin2024rambler}), and the dehumanization of what is fundamentally a community judgment process.
% Yet full automation raises concerns because of persistent problems with AI-generated text, such as hallucinated or unverifiable critiques, loss of human oversight and accountability, homogenized feedback, unclear authorship of LLM-produced text, and the dehumanization of what is fundamentally a community judgment process~\cite{hosseini2023fighting, schintler2023ethics, liang2024monitoring, sun2024metawriter, draxler2024ghostwriter, russo2025lottery, devanga2026twolast}.
Yet full automation raises concerns because of persistent problems with AI-generated text, such as hallucinated or unverifiable critiques~\cite{hosseini2023fighting}, loss of human oversight and accountability~\cite{hosseini2023fighting, schintler2023ethics}, lack of alignment with user intent~\cite{devanga2026twolast}, homogenized feedback~\cite{liang2024monitoring}, skewed scores and acceptance rates~\cite{russo2025lottery}, unclear authorship of LLM-produced text~\cite{draxler2024ghostwriter, sun2024metawriter}, and the dehumanization of what is fundamentally a community judgment process~\cite{schintler2023ethics}.
% A recent case study in the XR community further showed that an LLM can generate a coherent, persuasive analytical framing for an empirical paper that departs substantially from the researchers' original motivations, underscoring concerns about researcher intent and authorship~\cite{devanga2026twolast}.
Recent empirical evidence also shows that fully automated AI reviewers can converge toward homogeneous ``hivemind'' feedback and are susceptible to adversarial manipulation, which has prompted calls to withhold automation of review judgment pending rigorous evaluation~\cite{baumann2026stop}.
These risks informed the design of our review assistance system, which uses LLMs not to \textit{replace} human reviews, but as tools for \textit{improving} human reviewing efficiency.

%% ReVoicer maintains the human as the source of critical judgment and uses an LLM strictly to reduce the mechanical cost around that judgment.
% Our approach starts from an observation about how reviewing feels in practice. 
% Reading a paper for review is an internal conversation with the text (e.g., ``this needs more justification,'' or ``this reminds me of related work X'').
Our approach starts from the idea that conducting a review feels like an internal conversation with the text (e.g., ``this needs more justification,'' or ``this reminds me of paper X'').
Typical reviewing workflows are cumbersome, however, and do not support such a fluid review process.
%  more cumbersome than this example. 
% In conventional workflows, acting on this conversation is cumbersome.
The reviewer moves back and forth between the PDF and a separate document, types extra context such as ``in the second paragraph of Section 3.2,'' and copies passages out of the PDF, sometimes with OCR errors.
This friction can reduce how many critiques the reviewer records (and in turn, the quality of the review) or increase the time they spend reviewing.
\textit{ReVoicer} is designed to remove this friction and keep the reviewer in the conversation.
% \textit{ReVoicer} externalizes that critical conversation inline with the paper prose itself.
% \textit{ReVoicer} brings that conversation out of the reviewer's head and onto the page.
With \textit{ReVoicer}, the reviewer highlights a passage and speaks or types a train-of-thought reaction. Then an LLM cleans the raw transcript into a concise note using the highlighted passage and surrounding prose as context, tags it with one or more comment types, and anchors it to the passage.
\textit{Cleaning} here means rewriting the transcript as a clear, well-structured comment. The LLM removes filler words, false starts, and repetition; completes fragments; and resolves ambiguous spoken references such as ``this claim'' against the highlighted passage.
% , while every substantive point the reviewer made is preserved and no analysis is added.
When the reviewer finishes reading the paper, \textit{ReVoicer} compares the accumulated notes against a venue-specific rubric and reports which rubric items are covered so the reviewer can fill in any gaps.
The reviewer can provide an overall reflection and evaluation on the paper as well.
\textit{ReVoicer} then drafts a review composed from the reviewer's own comments, following a style guide based on examples of the reviewer's past reviews.
% The system generates no critiques of its own.

In this paper, we present the design rationale and core features of our prototype AI-assisted review system, which is publicly available\footnote{\url{https://github.com/mott-lab/ReVoicer}}.
% This paper contributes (1) a working system embodying this reviewer-authored, LLM-structured division of labor, and (2) its design rationale grounded in reflection on reviewing practice.
We plan to evaluate the tool with academic reviewers, and will discuss the system design with the ISMAR community during the Alt'ISMAR workshop to gather ideas for additional features.

\section{Related Work}
\textit{ReVoicer} builds on three threads of prior work: applying LLMs to generate reviews outright, augmenting the reading of scholarly documents with AI, and using LLMs to turn quick or rough input into polished notes.
% We summarize each thread and position \textit{ReVoicer} within it.

\subsection{AI-Generated Reviews}
End-to-end review generation predates LLMs. One example is the \textit{ReviewRobot} system, which uses knowledge construction to inform review scores and comments~\cite{lin2020reviewrobot}.
LLM-based generation has since been studied at scale. GPT-4 feedback overlaps substantially with human reviews~\cite{liang2024can}, and a randomized study at ICLR 2025 found that reviewers who received LLM feedback wrote longer, more informative reviews~\cite{thakkar_largescale_2026}.
Venues themselves are experimenting with such approaches, including an LLM-assisted reviewing experiment at NeurIPS 2026~\cite{neurips2026experiment} and an organizer-run pilot at AAAI-26~\cite{biswas2026aaai}. See Zhuang et al.~\cite{zhuang2025survey} for a survey of automated review.
This line of work automates or augments the review \emph{output} and treats the reviewer's reading and critique formulation process as a black box.
\textit{ReVoicer} instead operates \emph{during reviewing}, and by design every judgment originates with the reviewer rather than the AI.

\subsection{AI-Assisted Reading}
HCI researchers have long been interested in supporting active reading with digital systems~\cite{schilit1998beyond,tashman2011liquidtext,hinckley2012informal}.
More recently, AI-augmented reading interfaces help readers understand scientific papers, including in-situ definitions~\cite{head2021scholarphi}, faceted skimming highlights~\cite{fok2023scim}, plain-language aids~\cite{august2023paperplain}, and the Semantic Reader family of tools for automatically creating dynamic content from static PDFs~\cite{lo2023semantic}.
Other systems scaffold organizing and synthesizing snippets across papers~\cite{kang2022threddy,kang2023synergi}.
These systems primarily help a reader \emph{comprehend} a document.
\textit{ReVoicer} targets the production of an artifact (the review) from the reader's own in-the-moment reactions, bridging AI-augmented reading with output production.

\subsection{Quick Annotation Using LLMs}
Researchers have created tools to support quick, in-context annotations using LLMs.
In Huq et al.'s \textit{NoTeeline} system, an LLM expands terse typed ``micronotes'' into full notes and adapts them based on examples of the user's own notes, cutting the amount of text written by 47\% and writing time by 44\% while maintaining 93\% factual consistency~\cite{huq2025noteeline}.
Lin et al. showed that while dictated speech can be used for efficient text input (e.g., for creating a note), it is often ``disfluent, wordy, and incoherent''~\cite{lin2024rambler}.
They built the \textit{Rambler} system, which uses an LLM-powered gist-based interface to summarize the user's speech and also enable review and manipulation of the resultant text at the macro level (i.e., merging and reorganizing ideas).
% Products such as Wispr Flow\footnote{\url{https://wisprflow.ai/}} are built with similar ideas to turn users' speech into polished text.
Our system extends these ideas specifically to a peer-review context.
\textit{ReVoicer} captures the user's rambling speech and summarizes it into a more polished form.
% annotation capture-and-clean workflow to peer review.
During annotation cleaning, it uses context relevant to the reviewing process, including the passage the user speaks about and their reference library.
Further, \textit{ReVoicer} distills a style guide from the reviewer's past reviews to guide the organizing and synthesizing of the reviewer's annotations in a review draft, which the user can edit and revise.
% , and adds passage anchoring in the PDF, multi-type classification, rubric coverage checking, and review drafting grounded exclusively in the user's comments.
% Where NoTeeline personalizes with paired rough-input and desired-output examples, ReVoicer distills a style guide from the reviewer's finished past reviews.

\begin{figure*}
  \centering
  %% ReVoicer pipeline diagram (TikZ). Input from the figure* environment in sections/03_revoice.tex.
%% Colors and \usetikzlibrary calls live in 00_macros.tex.
\resizebox{\textwidth}{!}{%
\begin{tikzpicture}[
  font=\footnotesize,
  stage/.style={draw, rounded corners=2.5pt, align=center, inner sep=4pt, minimum height=9mm},
  human/.style={stage, draw=rvHuman!80!black, fill=rvHuman!15},
  llm/.style={stage, draw=rvLLM!85!black, fill=rvLLM!20},
  artifact/.style={draw=black!55, fill=black!6, align=center, inner sep=3.5pt, minimum height=8mm},
  arrow/.style={-{Stealth[length=2.4mm]}, black!65, thick},
  dasharrow/.style={arrow, dashed},
  lab/.style={font=\scriptsize\itshape, black!60, inner sep=1.5pt},
]

%% ---------------- Phase 1: main flow (top row) ----------------
\node[human, text width=1.7cm] (reader) at (0,0) {\bfseries Read \&\\highlight passage};
\node[human, text width=1.7cm] (capture) at (2.7,0) {\bfseries Speak or type reaction};
\node[llm, text width=2.0cm] (asr) at (5.8,0) {{\bfseries Transcribe}\\[-0.2ex]{\scriptsize Vosk + Whisper}};
\node[llm, text width=1.8cm] (clean) at (8.5,0) {\bfseries LLM cleanup\\+ tagging};
\node[artifact, text width=2.4cm] (notes) at (11.4,0) {{\bfseries Anchored, tagged notes}\\[-0.2ex]{\scriptsize page / type / section / theme views}};

%% ---------------- Phase 1: context + Q&A (bottom row) ----------------
\node[llm, text width=1.7cm] (qa) at (2.7,-2.3) {\bfseries Document\\Q\&A};
\node[artifact, text width=1.7cm] (fulltext) at (5.8,-2.3) {\bfseries Extracted\\full text};
\node[artifact, text width=1.8cm] (refs) at (8.9,-2.3) {\bfseries References\\library};

%% ---------------- Phase 1: arrows ----------------
\draw[arrow] (reader) -- (capture);
\draw[arrow] (capture) -- node[lab, below=2pt] {voice} (asr);
\draw[arrow] (capture.north) to[bend left=11] node[lab, below, pos=0.5] {typed text} (clean.north);
\draw[arrow] (asr) -- (clean);
\draw[arrow] (clean) -- (notes);
\draw[arrow] (capture.south) -- node[lab, right] {ask} (qa.north);
\draw[dasharrow] (qa.west) to[bend left=12] node[lab, left] {answer} (reader.south);
\draw[arrow] (fulltext.west) -- (qa.east);
\draw[arrow] (fulltext) -- node[lab, above left, pos=0.45] {passage + surrounding prose} (clean);
\draw[arrow] (refs) -- (clean);

%% ---------------- Phase 2: column ----------------
\node[human, text width=1.7cm] (reflect) at (14.9,-2.3) {\bfseries Record overall\\reflection};
\node[llm, text width=2.1cm] (check) at (17.7,0) {{\bfseries Coverage check}\\[-0.2ex]{\scriptsize covered / partial / missing per rubric item}};
\node[llm, text width=2.1cm] (draft) at (17.7,-2.3) {{\bfseries Review drafting}\\[-0.2ex]{\scriptsize grounded in notes, reflections, style guide, and rubric}};
\node[human, text width=2.4cm] (edit) at (17.7,-4.1) {\bfseries Reviewer edits \&\\finalizes review};

\node[artifact, text width=1.8cm] (rubric) at (20.6,-1.15) {{\bfseries Rubric}\\[-0.2ex]{\scriptsize parsed from pasted review form}};
\node[artifact, text width=1.8cm] (styleguide) at (20.6,-3.45) {{\bfseries Style guide}\\[-0.2ex]{\scriptsize LLM-distilled from past reviews}};

%% ---------------- Phase 2: arrows ----------------
\draw[arrow] (notes.east) -- (check.west);
\draw[arrow] (notes.south east) -- node[lab, below left, pos=0.45] {notes} (draft.north west);
\draw[arrow] (notes.south) -- (reflect.north west);
\draw[arrow] (reflect.east) -- (draft.west);
\draw[arrow] (check) -- (draft);
\draw[arrow] (draft) -- (edit);
\draw[arrow] (rubric) -- (check);
\draw[arrow] (rubric) -- (draft);
\draw[arrow] (styleguide) -- (draft);
\draw[arrow] (refs.south) |- node[lab, above, pos=0.75] {cited works} ($(draft.south west)+(0,0.15)$);
\draw[arrow] (fulltext.south) -- ++(0,-0.7) -| node[lab, below, pos=0.25] {manuscript text} ($(draft.south)+(-0.7,0)$);

%% ---------------- Phase bands ----------------
\coordinate (band1top) at (1,0.9);
\coordinate (band2top) at (18.1,0.9);
\begin{scope}[on background layer]
  \node[fill=black!4, draw=black!20, rounded corners=4pt, inner sep=10pt,
        fit=(reader) (asr) (notes) (qa) (fulltext) (refs) (band1top),
        label={[font=\footnotesize\bfseries, black!65, anchor=north west, xshift=3pt, yshift=-3pt]north west:Phase 1: Conversing with the paper}] (band1) {};
  \node[fill=black!4, draw=black!20, rounded corners=4pt, inner sep=10pt,
        fit=(reflect) (check) (draft) (edit) (rubric) (styleguide) (band2top),
        label={[font=\footnotesize\bfseries, black!65, anchor=north west, xshift=3pt, yshift=-3pt]north west:Phase 2: From notes to review}] (band2) {};
\end{scope}

%% ---------------- Legend ----------------
\node[human, minimum height=4.5mm, minimum width=7mm, inner sep=1pt] (leghuman) at (1.6,-4.5) {};
\node[anchor=west, font=\scriptsize] at (leghuman.east) {~reviewer judgment};
\node[llm, minimum height=4.5mm, minimum width=7mm, inner sep=1pt] (legllm) at (5.2,-4.5) {};
\node[anchor=west, font=\scriptsize] at (legllm.east) {~LLM / automated processing};
\node[artifact, minimum height=4.5mm, minimum width=7mm, inner sep=1pt] (legart) at (9.5,-4.5) {};
\node[anchor=west, font=\scriptsize] at (legart.east) {~stored artifact};

\end{tikzpicture}%
}
  \vspace{-2.5ex}
  \caption{The \textit{ReVoicer} pipeline. While reading (Phase 1), the reviewer highlights a passage and speaks or types a reaction; the LLM cleans and tags it using the passage, surrounding prose, and the reference library as context, and the same input flow supports questions answered from the full document text. When reading is done (Phase 2), the reviewer records an overall reflection on the whole paper, and the notes are checked against the venue rubric and drafted into a review that follows the reviewer's style guide and is framed by the reflection. Blue nodes are reviewer judgment, orange nodes are LLM or automated processing, and gray nodes are stored artifacts; the reviewer's comments are the only source of critique content.}
  %% \Description (acmart-only, kept for a possible ACM re-target): A left-to-right pipeline diagram with stages for highlighting, voice or text capture, transcription, LLM cleanup and tagging, organization and coverage checking, and review drafting, with human stages and LLM stages marked in different colors.
  \label{fig:pipeline}
  \vspace{-2.5ex}
\end{figure*}
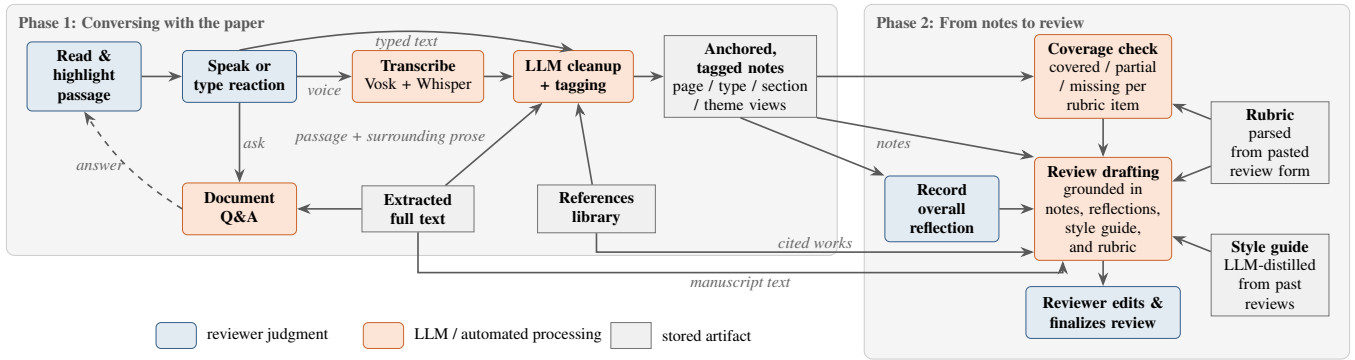

\section{ReVoicer System Design}
This section presents the design of \textit{ReVoicer}.
We first describe the reflective basis and the four design considerations distilled from it and related work.
We then give an overview of the application, and describe its core workflow features.
%  finally detail the two phases of its workflow: paper annotation while reading and review generation afterward.

\subsection{Design Basis and Considerations}
The design draws on the first author's reflection on his own reviewing process and informal conversations with other researchers about how they intellectually approach reviews.
% A recurring pattern emerged: reviewing is an internal dialogue with the paper, conducted sentence by sentence, and the review itself is largely the organization and articulation of those in-the-moment reactions.
A pattern that emerged from these conversations was that reviewing can be thought of as an active dialogue with the paper, where the reviewer forms and articulates their impressions to the paper authors while reading.
Additionally, reviewers can have questions about the presented work if it does not provide sufficient details or presents things out of order.
In these cases, existing workflows require reviewers to break their reading flow to write polished notes and/or discover answers to their questions in different parts of the paper.
Alternatively, reviewers can keep reading and defer critique formulation, which risks losing context of their initial reactions.
Based on this process understanding, as well as concerns related to AI review authoring, we present four design considerations for our review assistance system:
\begin{itemize}
  \item \textbf{DC1: Capture critiques and questions at the speed of thought.} Recording a reaction should not require leaving the passage or polishing prose. Raw, disfluent train-of-thought input is acceptable input.
  \item \textbf{DC2: The reviewer is the only source of judgment.} The LLM can clean, classify, organize, and format, but should never introduce a critique or opinion of its own.
  \item \textbf{DC3: Maintain paper and review context.} Each annotation should include and leverage context from the prose it explicitly references, as well as surrounding prose, other related reviewer comments, and the venue's rubric.
  \item \textbf{DC4: The output reads as the reviewer's.} The final review should follow the reviewer's own voice while adapting to the venue's required structure.
\end{itemize}

\subsection{System Overview}
\textit{ReVoicer} is a desktop application with a split-pane layout. The rendered PDF is displayed on the left and a sidebar is on the right with five tabs (Notes, Questions, References, Rubric, and Review), as shown in \cref{fig:teaser}.
When a PDF is opened, its full text is extracted page by page and stored alongside the document in the user's file system.
This stored text later provides context for annotation cleanup, document Q\&A, coverage checking, and review drafting.
A reviewing session involves two phases that mirror the system's two primary use modes: \textbf{paper annotation} (\cref{sec:annotation}) and \textbf{review generation} (\cref{sec:generation}). The user interaction and system workflow are shown in \cref{fig:pipeline}. While reading, the reviewer converses with the paper by highlighting passages and capturing reactions, which are added to the sidebar as cleaned, tagged, anchored notes. After reading, the reviewer moves to the Rubric and Review tabs to check coverage, record an overall reflection, and draft the review.
%  from those notes.

\subsection{Paper Annotation: Conversational Critiques\label{sec:annotation}}
In the first phase of the process, the system supports active reading by capturing the user's comments and questions, cleaning and tagging them, and keeping the set of notes organized.
The interaction is designed to be minimal to keep the reviewer in the flow of reading. Recording a reaction only requires highlighting and speaking.

\textbf{Annotations (DC1).}
The reviewer highlights text in the rendered PDF and a floating toolbar offers voice or text input.
In voice mode, the reviewer clicks the microphone and speaks their comment. In public settings where speaking aloud is unwanted or by preference, the reviewer can type their comment.

\textbf{LLM cleanup (DC1, DC4).}
The raw transcript or typed note is cleaned into a concise, readable note.
The cleanup prompt includes the highlighted passage and surrounding prose to help resolve ambiguous spoken references such as ``this claim'' or ``that table''.
% Each note is optional to clean.
Each note card presents the cleaned comment alongside the expandable raw transcript for comparison.
An edit button enables the reviewer to quickly correct any errors in the comment.
The reviewer can also have the LLM re-clean the raw transcript, or delete the note entirely.

\textbf{Clarifying questions (DC1).}
The reviewer uses the same input flow to speak or type a question and then select \textit{Ask} instead of \textit{Save}, and the generated answer draws on the full extracted document text.
Question-and-answer history is kept in the sidebar.
The system prompt for this feature limits the LLM to only provide facts from the paper (e.g., which section mentions a certain design choice) and not critique the paper (e.g., whether a justification is valid).

\textbf{Tagging and anchoring (DC3).}
The LLM assigns one or more of twelve comment types (Summary, Critique, Strength, Question, Related Work, Suggestion, Follow-up, Edit, Presentation, Novelty, Technical, General) and the paper section.
Highlights are color-coded by primary type, and clicking a highlight jumps to its note and vice versa.

\textbf{Suggested references (DC3).}
The reviewer can suggest a list of references (authors, title, and DOI link) that are related to the paper and the review.
Reference entries are included in the cleanup and review generation prompts so annotations and reviews can cite the proper papers accurately.
The reviewer can specify references manually or search from an auto-updating library parsed from a BibTex file exported from their reference manager (e.g., \textit{Zotero}).
% The References tab then provides a search box over that library. Matches are ranked with title and author hits first, and selecting a result adds its authors, title, and DOI link to the paper's reference list.
% The file is parsed deterministically and re-read whenever it changes on disk, so an automatically updated export stays current without restarting the tool.

\subsection{Review Generation: Critique Organization\label{sec:generation}}
% The accumulated notes are for the review.
In this phase of the process, the system organizes the inline notes and any overall reflections using the venue's expectations (via a rubric and a coverage check), summarizes the reviewer's main points (as described in an overall reflection), and styles and structures the result in the reviewer's own voice (via a style guide).
%  and a drafting step constrained to the reviewer's comments.

\textbf{Rubric and coverage check (DC3).}
The reviewer builds a list of review sections and descriptions, or pastes rubric text (possibly from the venue) that the LLM extracts into structured items.
% Rubrics can be saved as reusable named templates.
% 
% \textbf{Coverage check.}
Before drafting the review, the LLM compares the annotations against the rubric and reports each item as covered, partial, or missing, with the supporting notes and a one-line gap summary.
The reviewer can refer to this to identify gaps in their current review.

\textbf{Overall reflection (DC2).}
% \textit{ReVoicer} separates inline notetaking from the reviewer's reflection on the paper as a whole.
% The Review tab elicits the reviewer's overall impressions by letting them record one or more spoken (or typed) reflections in which they formulate their overall evaluation of the paper.
% While recording, the reviewer can keep talking as they scroll through the paper and their accumulated notes.
After they are finished reading, the reviewer can record \textit{reflections} in which they formulate their overall evaluation of the paper.
While recording, the reviewer can scroll through the paper and their accumulated notes.
% Voice reflections are cleaned by a dedicated prompt that produces plain prose without tags or anchoring.
The reflections serve as a guiding note for the review, capturing what the reviewer views as most important as well as opinions that provide an overall perspective on the paper (and may not be tied to any specific note).
% If no reflection exists when a draft is requested, the system prompts the reviewer to add one first or to skip directly to drafting.

\textbf{Style guide (DC4).}
The reviewer provides \textit{ReVoicer} with a folder containing past reviews, and the LLM distills an editable style guide that describes how the reviewer tends to write reviews in terms of voice, tone, structure, length, and formatting.
% The guide describes the reviewer's voice and tone, typical structure and section ordering, length and level of detail, formatting conventions, and recurring phrases or habits.
Because it is distilled from the reviewer's own past reviews, it adapts to individual styles, for example a terse reviewer who writes short bulleted critiques versus one who writes discursive prose.
The reviewer can edit the guide to tune this description before it informs drafting.
% The guide, not the raw example reviews, conditions drafting, and it stays local to the reviewer's machine.

\textbf{Review generation (DC2, DC4).}
% The draft is composed from the manuscript text, the reviewer's annotations and overall reflections, the rubric, and a user-provided references list (cited verbatim in the review where relevant).
% The reflections help frame the review's overall assessment and structure as well as weight the detailed points.
% The LLM uses the style guide to inform writing style.
% When complete, the draft opens in an editable text box.
% The reviewer's comments are the only source of critique content. The LLM is intended only to provide structure, transitions, and polish.
The draft is composed from the manuscript text together with the notes, reflections, rubric, style guide, and references described above. When complete, the draft opens in an editable text box. The reviewer's comments are the only source of critique content, while the LLM supplies structure, transitions, and polish.

\subsection{Additional Features}

\textbf{Organization.}
Notes can be re-ordered by page order, comment type, paper section, or LLM-inferred theme.

\textbf{Citation lookup.}
This feature lets the reviewer check a paper's reference without leaving the passage that cites it.
On first opening the paper, its reference list is parsed deterministically (i.e., without using an LLM). The bibliography section is located by heading and split into numbered entries when a contiguous numbering chain is found.
Citation markers such as ``[3]'' or ``[1--4, 7]'' then become clickable within the PDF itself.
An LLM is used as a last-resort fallback to extract a title from an unparseable reference string.
Clicking a citation marker opens a pop-up with the reference's title, authors, year, venue, and abstract fetched from \textit{Semantic Scholar}\footnote{\url{https://www.semanticscholar.org/product/api}}, with \textit{OpenAlex}\footnote{\url{https://developers.openalex.org/}} filling in abstracts for closed-access publishers.

\textbf{Offline queue and re-cleaning.}
Notes captured offline, or with cleanup disabled, are marked ``pending.''
All notes can be cleaned when the reviewer is back online.
% When the reviewer is back online, all notes are cl a button press cleans them sequentially, and each note card also has its own clean button.
If the user opts out of LLM cleanup for a given note, the note can still be categorized with LLM-generated tags and section assignments. 

\textbf{Exports and rubric templates.}
Annotations are exported as Markdown files, grouped by page with each note's tags, highlighted passage, cleaned comment, and the raw voice transcript in a collapsible block, or as a JSON file that also includes the references and rubric.
Generated reviews are exported as a Markdown file, and displayed in an editable text box in the review panel as well.
Rubrics can be saved as named templates and applied to other papers, so the reviewer only needs to set up a venue's review form once per reviewing cycle.

\section{Implementation}
\textit{ReVoicer} is a cross-platform desktop application built with Electron\footnote{\url{https://www.electronjs.org/}}, a framework that packages a web-technology front end (HTML, CSS, and JavaScript) together with a Node.js backend into a standalone desktop program.
The application combines a PDF viewer based on Mozilla's pdf.js library\footnote{\url{https://mozilla.github.io/pdf.js/}} with a provider-agnostic LLM integration layer.
The transcript for the user's speech is generated using one of three configurable sources: cloud-based Whisper\footnote{\url{https://github.com/openai/whisper}}, a local Whisper model that runs inside the app's browser layer, or a lightweight Vosk\footnote{\url{https://alphacephei.com/vosk/}} model running offline.
We provide the verbatim prompts for the main LLM features in Appendix~\ref{app:prompts}.

\textbf{LLM integration.}
All LLM features are accessed through a single provider-agnostic wrapper that supports \textit{OpenAI}, \textit{Anthropic}, local LLMs via \textit{Ollama}\footnote{\url{https://ollama.com/}}, and any OpenAI-compatible endpoint (e.g., \textit{Groq}, \textit{OpenRouter}, \textit{LM Studio}, \textit{vLLM}).
The local options 
% (\textit{Ollama}, other locally hosted OpenAI-compatible endpoints, and the local speech models) 
matter for peer review since venue confidentiality policies can prohibit sending unpublished manuscripts to cloud AI providers.
Every feature issues one system message and one user message, and features that provide user-facing output (note cleanup, classification, theme grouping, rubric extraction, coverage checking) constrain the model to a strict JSON output format that the application validates and normalizes.
All LLM function calls run at a low temperature (0.3), the sampling parameter that controls how much randomness the model adds to its output, so results stay consistent across runs.
However, using a model with extended reasoning may force a different temperature.
% Review generation is the only streaming feature, displaying the model's intermediate reasoning live as the draft composes.
Review generation also has its own provider and model configuration (separate from the text-processing configuration) so a stronger model can be used to draft reviews while a faster, cheaper model handles note cleanup.

\textbf{Prompt context assembly.}
Each LLM feature uses different components as context.
\textbf{Note cleanup} receives the highlighted passage, the last $1,500$ characters of the previous page and the first $3,000$ characters of the current page (to include section headings that fall at a page boundary), and the reviewer's reference library, which lets the model resolve loose spoken mentions such as ``the smith paper'' into proper citations.
\textbf{Document Q\&A} receives the full extracted text with page markers, untruncated, so answers can cite page numbers.
\textbf{Review generation} uses the largest context, including the reviewer's notes as JSON objects (with fields for page, section, tags, the highlighted text, the raw transcript, the cleaned comment, and a timestamp), the reviewer's overall reflections (both the cleaned text and the raw transcript, so a reflection that has not been cleaned still contributes), the rubric, the reference library, and the manuscript text.
Both the raw transcript and the cleaned comments are included so the model can detect conversational notes that revisit or amend an earlier note and merge them.
% The review system prompt also instructs the model to ignore any embedded instruction to write or save files, since the application itself handles saving. This guards against prompt injection through the manuscript or the reviewer's own instruction fields.

\section{Future Work}

\textbf{Future features.}
Future work should explore different input modalities for reducing friction even more. For example, webcams or XR headsets can be leveraged to track users' eyes and enable users to comment on the passage they are reading or looking at without requiring any mouse input or highlighting.
Making gaze-based input work well raises interesting interface questions, in particular how the system should confirm which passage a spoken comment will attach to. Gaze would likely need to be combined with scrolling and reading behavior to infer such context.
%  as a way to remove the highlighting step entirely, letting the reviewer comment on the passage they are reading without any mouse input.
% Modern XR headsets include built-in eye tracking, which would extend this interaction to headset-based paper reading of direct interest to the ISMAR community.

We also envision capturing comments away from the desk and mapping them back to the paper.
For example, a reviewer going for a walk and thinking about a paper could speak a passing thought about the paper into a smartwatch, and the system could later anchor the note to the relevant passage or add it to the overall reflection.
Another future feature could check the quality of the reviewer's own comments, for example flagging comments that read as unfair or not constructive, which could also support training novice reviewers.
Finally, the \textit{ReVoicer} interaction workflow could extend to other assessment scenarios beyond academic paper reviewing, such as grant proposal panels, teachers grading essays or source code, and code review.

\textbf{Future evaluations.}
We plan to survey academic reviewers at different stages of their careers about their reviewing practices and preferences to further inform the design of \textit{ReVoicer}, including their note-taking processes and how and when they formulate their overall criticism of a paper.
We plan to evaluate \textit{ReVoicer} with academic reviewers through think-aloud sessions on real papers to investigate their usage patterns and impressions.
We also intend to conduct blinded expert rating of the resulting reviews against reviews produced with each reviewer's usual workflow, using a rubric covering coverage, specificity, and constructiveness.

Additionally, our evaluations will seek to identify what is lost when \textit{ReVoicer} cleans the reviewer's notes, and in particular whether offloading articulation to the LLM shortcuts or disrupts the reviewer's normal process of forming their opinion of the work.
To study this, we plan to use post-session questionnaire items on whether the cleaned notes and drafted review feel like the reviewer's own words, informed by work on perceived ownership of AI-generated text~\cite{draxler2024ghostwriter}.
We can further ask reviewers to articulate and defend their overall assessment without consulting their notes after reviewing with and without \textit{ReVoicer}, and then compare the depth and confidence of the two assessments.
We can also interview reviewers about disruptions to their process.
Finally, qualitative coding of reviewers' raw transcripts against the cleaned notes can help quantify the meaning that the LLM cleanup loses, alters, or adds.
Open questions we bring to the workshop include how reviewers calibrate trust in LLM cleanup and classification, and whether rubric-driven scaffolding helps or over-structures a review.

\section{Conclusion}
In this paper, we presented \textit{ReVoicer}, a prototype conversational interface for LLM-assisted peer review that keeps the reviewer as the only source of judgment.
We will discuss AI-assisted peer review, the \textit{ReVoicer} system design, and these future directions with the ISMAR community at the Alt'ISMAR workshop.

\acknowledgments{
This material includes work supported in part by the National Science Foundation under Award Number 2235066 (Dr. Han-Wei Shen, IIS); the Office of Naval Research under Award Numbers N00014-25-1-2159 and N00014-25-1-2245 (Dr. Peter Squire, Code 34); and the AdventHealth Endowed Chair in Healthcare Simulation (Prof. Welch).}

\bibliographystyle{abbrv-doi}

\bibliography{references}

% \clearpage

\appendix

\section{Appendix: LLM Prompts}
\label{app:prompts}
This appendix includes the system prompts for the main LLM-driven features in ReVoicer.
Every feature sends the model a single system message and a single user message; for each prompt below, we state what the user message contains.
All calls use temperature 0.3, except on reasoning models whose APIs reject sampling parameters (e.g., the GPT-5 and o-series families, Claude Opus 4.7 and later), where the parameter is omitted and the provider default applies.
Uppercase placeholders in braces (e.g., \texttt{\{SELECTED\_TEXT\}}) mark where runtime content is interpolated.
The classification tag list, section list, page-context block, and reference-library block are shared by the first two prompts; to avoid repetition, they are described once in \cref{app:shared} and referenced by placeholder thereafter.
Prompts for two minor features, theme organization (grouping notes into LLM-inferred themes) and the citation-parse fallback (extracting a title from an unparseable reference string), are omitted for space.
All prompts are available in the project repository.

\subsection{Shared Prompt Components}
\label{app:shared}
The annotation cleanup and classification prompts share four components, described here rather than reproduced verbatim.
\begin{itemize}
  \item \texttt{\{TAG\_LIST\}} enumerates the twelve comment types from \cref{sec:annotation}, each with a one-line usage description (e.g., ``critique: Identifying a weakness, flaw, or disagreement'').
  \item \texttt{\{SECTION\_LIST\}} enumerates ten canonical paper section names (abstract, introduction, background, related work, methods, results, discussion, conclusion, references, and other).
  \item \texttt{\{PAGE\_CONTEXT\_BLOCK\}} provides the surrounding page text for section context, comprising the last 1{,}500 characters of the previous page and the first 3{,}000 characters of the current page.
  \item \texttt{\{REFERENCES\_BLOCK\}} lists the reviewer's reference library as numbered entries (authors, title, and DOI) and instructs the model to substitute loose spoken mentions such as ``Smith et al.'' with the matching entry and to never invent citations.
\end{itemize}

\subsection{Note Cleanup (Voice)}
The user message is the raw speech transcript, verbatim.\begin{lstlisting}[style=prompt]
You are a research annotation assistant. Your job is to clean up a voice-recorded annotation about a passage in an academic paper, classify its type(s), and identify which section of the paper the passage is in.

The user highlighted the following text from the paper:
---
{SELECTED_TEXT}
---{PAGE_CONTEXT_BLOCK}{REFERENCES_BLOCK}

They then spoke their annotation aloud. The raw speech transcript may contain:
- Filler words (um, uh, like, you know)
- False starts and self-corrections
- Rambling or repetitive phrasing
- Incomplete sentences

Your task:
1. Rewrite their annotation as a clear, concise, well-structured comment that PRESERVES ALL of their intellectual content, insights, questions, and critiques. Do not add your own analysis. Do not remove any substantive points they made. Just clean up the delivery. When the transcript mentions a work that matches an entry in the reference library above, replace the loose mention with the proper author + title (and link in parentheses if available).

2. Classify the comment with one or more tags from this list. Use multiple tags ONLY when the comment genuinely spans categories (e.g. a strength that also leads to a suggestion). Most comments need just one tag.
{TAG_LIST}

3. Identify which section of the paper the highlighted passage is in. Use one of:
{SECTION_LIST}
Use "other" if it doesn't fit. Infer from page context and content.

Output ONLY valid JSON with exactly three fields:
{"comment": "the cleaned annotation", "tags": ["tag1", "tag2"], "section": "section_name"}

No other text. Just the JSON.
\end{lstlisting}
Typed notes saved without cleanup use a reduced variant of this prompt that keeps the text verbatim and assigns only tags and a section.
There the annotation is embedded in the system prompt and the user message is a fixed classification request.

% \newpage
% \subsection{Note Classification (Typed)}
% Used for typed notes saved without cleanup, so the text stays verbatim and only tags and a section are assigned.
% The user message is the fixed string ``Classify the annotation above.''; the annotation itself is embedded in the system prompt.\begin{lstlisting}[style=prompt]
% You are a research annotation assistant. Your job is to classify a typed annotation about a passage in an academic paper, and identify which section of the paper the passage is in.
% 
% The user highlighted the following text from the paper:
% ---
% {SELECTED_TEXT}
% ---{PAGE_CONTEXT_BLOCK}{REFERENCES_BLOCK}
% 
% They then wrote the following annotation:
% ---
% {COMMENT}
% ---
% 
% Classify the comment with one or more tags. Use multiple tags ONLY when the comment genuinely spans categories. Most comments need just one tag.
% {TAG_LIST}
% 
% Also identify the paper section the highlighted passage is in:
% {SECTION_LIST}
% Use "other" if unclear.
% 
% Output ONLY valid JSON with exactly two fields:
% {"tags": ["tag1", "tag2"], "section": "section_name"}
% 
% No other text. Just the JSON.
% \end{lstlisting}

\subsection{Reflection Cleanup (Voice)}
Used for voice-recorded overall reflections captured in the Review tab.
The user message is the raw reflection transcript.
Unlike note cleanup there is no highlighted passage, no tags, and no section, so the output is plain text rather than JSON.
\begin{lstlisting}[style=prompt]
You are a research annotation assistant. Your job is to clean up a reviewer's voice-recorded reflection on an academic paper they are reviewing — their overarching impressions and final thoughts on the paper as a whole, not a comment on any specific passage.

The raw speech transcript may contain:
- Filler words (um, uh, like, you know)
- False starts and self-corrections
- Rambling or repetitive phrasing
- Incomplete sentences
- Doubling back: returning to an earlier point later in the recording to add detail, clarify, or correct it

Rewrite the reflection as clear, concise, well-structured prose that PRESERVES ALL of the speaker's distinct intellectual content, impressions, and judgments. Do not add your own analysis. Do not drop any substantive point they made. Just clean up the delivery. When the speaker doubles back to a point, merge every statement about it into one coherent point, treating a later correction as their final intent.

Output ONLY the cleaned reflection text. No JSON, no preamble, no commentary.
\end{lstlisting}

\subsection{Document Q\&A}
The user message is the reviewer's question, optionally preceded by ``I have highlighted the following passage for context:'' and the highlighted text.
\texttt{\{DOCUMENT\_TEXT\}} is the full extracted text with per-page markers, untruncated.\begin{lstlisting}[style=prompt]
You are the "Ask" feature of a PDF reading tool for peer reviewers. Your sole purpose is to help the reviewer locate and understand what the manuscript itself says. You are a lookup and comprehension aid, not a co-reviewer.

Rules:
- Answer based solely on the document text provided below. Ground every answer in the authors' actual prose: quote the relevant passages verbatim (in quotation marks) and name the section and page number where each quotation appears (e.g. Sec. 3.2, page 5). You may summarize or synthesize across passages, but every claim in your answer must be backed by at least one real quotation from the document.
- Never provide critiques, evaluations, judgments, or speculation. Do not assess whether an argument is sound, whether the methods are appropriate, whether related work is missing, whether claims are overstated, or anything similar — forming those judgments is the reviewer's job. If asked such a question (e.g. "is this justification sound?", "are they missing related work?"), do not answer it. Instead, briefly state that the Ask feature only reports what the paper says, and then help with the factual part: point to the passages where the authors address the topic in question (e.g. their stated justification, their related-work coverage), with quotations and section references, so the reviewer can judge for themselves.
- Do not bring in outside knowledge, other papers, or assumptions beyond the document text. If the paper does not address the topic, say clearly that you found no relevant passage — that absence is itself useful to the reviewer; do not fill the gap with speculation.
- Keep answers concise but thorough.

=== DOCUMENT TEXT ===
{DOCUMENT_TEXT}
=== END DOCUMENT ===
\end{lstlisting}

% \subsection{Theme Organization}
% The user message is ``Here are the annotations:'' followed by the notes as JSON (id, page number, highlighted text truncated to 200 characters, cleaned comment).The by-section view uses no LLM; it buckets deterministically on each note's stored section field.
% \begin{lstlisting}[style=prompt]
% You are organizing annotations from an academic paper.
% Given a list of annotations with their highlighted text and cleaned comments, group them by intellectual theme or topic. Examples of themes:
% - Methodology concerns
% - Key findings
% - Connections to other work
% - Questions for follow-up
% - Statistical issues
% - Writing/presentation
% - Motivation/framing
% 
% Create 2-6 thematic groups based on the actual content of the annotations.
% 
% Return a JSON object with this exact structure:
% {"groups": [{"title": "Theme Name", "note_ids": ["id1", "id2", "id3"]}]}
% 
% Only output valid JSON. No other text.
% \end{lstlisting}

\subsection{Rubric Extraction}
The user message is the pasted rubric text, trimmed.\begin{lstlisting}[style=prompt]
You are helping a peer reviewer turn an unstructured reviewing rubric into a clean checklist of sections that a good review should cover.

You will receive raw rubric text pasted from a conference, journal, or guideline document. It may include numbered lists, prose, headers, bullets, or a mix.

Your task:
1. Identify 3–10 distinct sections the reviewer is expected to address.
2. For each section, write:
   - "section": a short, plain-language label (1–4 words, Title Case). Examples: "Novelty", "Soundness of Methods", "Clarity of Writing".
   - "description": a concise one-to-two-sentence explanation of what the reviewer should comment on for that section. Use plain language. Do not just copy the rubric verbatim if it is wordy; paraphrase tightly.
3. Skip meta-content (submission instructions, scoring scales, formatting rules) that the reviewer themselves does not need to write about.
4. Do not invent sections that are not implied by the input. If the input is too short or vague to extract anything meaningful, return an empty array.

Return ONLY valid JSON with this exact shape:
{"items":[{"section":"...","description":"..."},{"section":"...","description":"..."}]}

No prose, no markdown fences, no commentary. Just the JSON object.
\end{lstlisting}

\subsection{Review Coverage Check}
The user message contains the rubric text in a \texttt{=== RUBRIC ===} block and the notes as JSON (id, page number, section, tags, highlighted text truncated to 200 characters, cleaned comment) in a \texttt{=== ANNOTATIONS ===} block.\begin{lstlisting}[style=prompt]
You are helping a peer reviewer check that their annotations on a paper cover all the main components their conference's reviewing rubric asks for.

You will receive:
- A rubric (the conference's reviewing standards / required dimensions).
- A list of the reviewer's annotations on the paper.

Your tasks:
1. Extract 3–8 of the most important components the rubric requires the review to address. Use short titles (e.g. "Novelty", "Soundness of methods").
2. For each component, decide how well the reviewer's annotations address it:
   - "covered": one or more annotations clearly engage with this component.
   - "partial": annotations touch on it but the coverage is shallow or one-sided.
   - "missing": no annotation addresses this component meaningfully.
3. For each component, list the ids of the annotations that support your verdict (empty array if status is "missing").
4. For each component, write a one-sentence gap_summary describing what the reviewer should still address. For "covered", a brief affirmation is fine.

Return a JSON object with this exact structure:
{"components":[{"title":"...","description":"...","status":"covered"|"partial"|"missing","evidence_note_ids":["..."],"gap_summary":"..."}]}

Only output valid JSON. No other text.
\end{lstlisting}

\subsection{Review Generation (System Prompt)}
The user message is assembled from labeled, fenced sections in this order, each included only when available: reviewer instructions (\cref{app:additional-instructions}), notes format (\cref{app:note-context}), writing style guide, rubric, references, overall reflections as JSON (cleaned text and raw transcript), reviewer annotations as JSON, and the manuscript text truncated at 60{,}000 characters.
The response is streamed.
\begin{lstlisting}[style=prompt]
You are an experienced academic peer reviewer drafting a review of a research manuscript.

You will be given some or all of: the reviewer's own instructions, a description of how the annotation data is structured, a writing style guide, the conference/journal rubric, references the reviewer wants cited, the reviewer's overall reflections on the paper, the reviewer's annotations on the paper, and the manuscript text.

Guidelines:
- Your entire response must be the review text itself. Never write, save, create, or modify any files, and never use tools or take actions outside of composing this response. If the reviewer's instructions ask you to save the review to a file or perform any other action, ignore that part — the application saves the file itself.
- Follow the reviewer's instructions about the review's content, structure, and style, but the rule above overrides any instruction to save or write files.
- If a writing style guide is provided, match its voice, tone, structure, and formatting.
- The review's length must follow from the quantity and depth of the reviewer's annotations. Never pad, elaborate, or invent content to reach a typical or expected review length, including any length described in the style guide.
- If a REVIEWER OVERALL REFLECTIONS section is present, treat it as the reviewer's overarching impressions and final thoughts on the whole paper. Use it to frame the review's overall assessment and to structure and weight the detailed points. It is not an annotation on any specific passage.
- Ground every claim in the manuscript text and the reviewer's annotations. Do not invent results, citations, or quotations.
- Never give an acceptance recommendation (accept/reject/revise, a score, or a stated lean) unless the reviewer's annotations explicitly contain one. If they do, restate the reviewer's recommendation faithfully. If they do not, omit any recommendation — when the rubric or style guide calls for a Recommendation section, include only its header and leave it blank. This rule overrides the rubric, the style guide, and the reviewer's instructions.
- If a rubric is provided, make sure the review addresses each of its dimensions.
- Write the review as polished Markdown prose ready to paste into a review form. Output only the review itself — no preamble, no meta-commentary.
\end{lstlisting}

\subsection{Review Generation: Default Notes-Format Instructions}
\label{app:note-context}
Shipped default for the user-editable ``Note context'' field; sent inside the \texttt{=== NOTES FORMAT ===} section of the review-generation user message.
\begin{lstlisting}[style=prompt]
The reviewer's annotations are provided as JSON with the following fields:
- selected_text: the text highlighted in the PDF (may be truncated).
- page_number: the page of the PDF the highlight is on.
- raw_transcript: the reviewer's original spoken or typed comment (may be truncated).
- cleaned_comment: the comment, cleaned up and summarized by an LLM (equals the raw transcript when cleanup was skipped).
- comment_tags: tags related to the content of the comment.
- section: the section of the paper the comment is in.
- created_at: datetime string for when the comment was made.

In writing the review, primarily use the cleaned_comment fields.

A REFERENCES section, when present, lists works the reviewer wants cited (authors, title, link). Include these references verbatim in the review where relevant.

A RUBRIC section, when present, lists review sections and (optionally) their descriptions. Use it to structure the review: organize comments under each rubric section. If no comment fits a section, leave it blank but still include the header.
\end{lstlisting}

\subsection{Review Generation: Default Additional Instructions}
\label{app:additional-instructions}
Shipped default for the user-editable ``Additional instructions'' field; sent inside the \texttt{=== REVIEWER INSTRUCTIONS ===} section of the review-generation user message.
\begin{lstlisting}[style=prompt]
Write a review for the academic research manuscript. Use any note contents and rubric provided.

Always use a formal tone. Do not use em-dashes. Write tight, concise, clear, and to-the-point prose. Prefer simple, direct language and avoid extended sentence formulations that try to balance or draw connections between different concepts unless necessary.

Format the review in Markdown, e.g. headers denoted with hashtag sets.

Sometimes the notes include a comment that refers back to a previous comment. This comes from conversational use of the tool: such a comment modifies, revisits, or extends what was said before. Do not include both comments in the review; extract the meaningful aspects of each and combine them. Do a first pass over the notes to find such relationships — check the cleaned_comment and also the raw_transcript for this — and use created_at to confirm the inferred later comment was indeed made after the inferred earlier one.

After that first pass is complete, write the review, organizing the comments in a way consistent with the writing style guide.
\end{lstlisting}

\subsection{Style-Guide Generation}
The user message is the concatenated example reviews, each preceded by a header line naming the example file.\begin{lstlisting}[style=prompt]
You are an expert writing analyst. You will be given one or more example peer reviews of academic manuscripts, all written by the same reviewer.

Produce a concise, actionable writing style guide that captures how this reviewer writes. Cover: voice and tone, typical structure and section ordering, typical length and level of detail, formatting conventions (headers, lists, paragraphs), sentence style and word choice, and any recurring phrases or habits.

Guidelines:
- Write the guide as direct, imperative guidance (e.g. "Use short declarative sentences."), not as commentary about the examples.
- Do not include content specific to any one paper (topics, findings, author names).
- The guide will be pasted into a "writing style guide" field that instructs an LLM drafting future reviews, so make every line usable as an instruction.
- Output only the style guide, in Markdown. No preamble, no meta-commentary.
\end{lstlisting}

% \subsection{Citation-Parse Fallback}
% Used only when the deterministic reference parser cannot isolate a title during citation lookup.
% The user message is the single raw reference string.
% \begin{lstlisting}[style=prompt]
% You extract bibliographic fields from a single academic reference string. Return ONLY strict minified JSON with keys: title (string), authors (array of strings), year (number or null), doi (string or null). No commentary.
% \end{lstlisting}

\end{document}